\documentclass[superscriptaddress,preprint,floatfix,amsmath,aps]{revtex4-1}
\usepackage{amsmath,amssymb,amsthm,latexsym,mathrsfs,amsfonts,}
\usepackage{graphicx,hyperref}
\usepackage{subfigure}
\usepackage{xcolor}
\usepackage{appendix}

\begin{document}

\title{Pressure-induced concomitant topological and metal-insulator quantum phase transitions in Ce$_3$Pd$_3$Bi$_4$}
\author{Chenchao Xu}
 \altaffiliation{Current Address: Center for Green Research on Energy and Environmental Materials (GREEN) and International Center for Materials Nanoarchitectonics (MANA), National Institute for Materials Science (NIMS) - Tsukuba, Ibaraki 305-0044, Japan}
 \affiliation{Department of Physics and Center for Correlated Matter, Zhejiang University, Hangzhou 310058, P. R. China} 

 \affiliation{Department of Applied Physics, University of Tokyo, Tokyo 113-8656, Japan}

 \author{Chao Cao}
  \email{ccao@zju.edu.cn}
 \affiliation{Department of Physics and Center for Correlated Matter, Zhejiang University, Hangzhou 310058, P. R. China} 
 \affiliation{Condensed Matter Group, School of Physics, Hangzhou Normal University, Hangzhou 311121, P. R. China}
  
  \author{Jian-Xin Zhu}
  \email{jxzhu@lanl.gov}
  \affiliation{Theoretical Division and Center for Integrated Nanotechnologies, Los Alamos National Laboratory, Los Alamos, New Mexico 87545, USA}
      
\date{\today}

\newpage
\thispagestyle{empty}

\begin{abstract}
The electronic property and magnetic susceptibility of Ce$_3$Pd$_3$Bi$_4$ were systemically investigated from 18 K to 290 K for varying values of cell-volume using dynamic mean-field theory coupled with density functional theory.  By extrapolating to zero temperature, the ground state of Ce$_3$Pd$_3$Bi$_4$ at ambient pressure is found to be a correlated semimetal due to insufficient hybridization.  Upon applying pressure, the hybridization strength increases and a crossover to Kondo insulator is observed at finite temperatures.  The characteristic temperature signaling the formation of Kondo singlet, as well as the characteristic temperature associated with $f$-electron delocalization-localization change,  simultaneously vanishes around a critical volume of 0.992$V_0$,  suggesting that such metal-insulator transition is possibly associated with a quantum critical point.  Finally, the Wilson's loop calculations indicate that the Kondo insulating side is topologically trivial, thus a topological transition also occurs across the quantum critical point.
\end{abstract}

\maketitle

\newpage
\thispagestyle{empty}

\newpage

\section{Introduction}
Kondo insulator is a prototypical strongly correlated quantum matter involving $4f$ or $5f$ electrons.
At higher temperatures, the system is metallic since the $f$-electron-derived  local moments do not strongly couple to conduction electrons.
Upon cooling, an energy gap opens at the Fermi energy, due to the formation of Kondo singlets between the local moments and conduction electrons.
In the meantime, the electronic states experience an incoherent to coherent crossover.
In the above process, the hybridization strength $\Gamma \propto  N_0 |V|^2$, with $V$ being the hybridization interaction and $N_0$ conduction electron density of states at the Fermi energy,  is of particular importance, since the Kondo coupling strength depends directly on this quantity as  $J_{\mathrm{K}} \propto |V|^2/U$ with $U$ the Coulomb interaction on $f$-electrons.
According to the famous Doniach phase diagram~\cite{doniach1977kondo},  this coupling leads to a conduction-electron mediated  Ruderman-Kittel-Kasuya-Yosida (RKKY) interaction, which in turns competes with the Kondo coherence effect.
They are characterized by respective energy scale, $T_{\mathrm{RKKY}}$ and $T_{\mathrm{K}}$.
Therefore, if $V$ is too small, the formation of Kondo singlet shall fail and lead to a magnetic ground state instead.
Apart from temperature, the strength of hybridization can be tuned via chemical doping, external pressure and magnetic field alternatively.
These nonthermal parameters therefore provide control over competing ground states, and may realize quantum critical phenomena~\cite{Sachdev:2008aa,QCP_RMP,gegenwart2008quantum}.
More interestingly, when the competing ground states are associated with different topological invariants, a topological quantum phase transition is achieved.
It has been proposed that the critical point of a topological quantum phase transition can host novel semimetal states, which exhibit non-Fermi liquid or marginal Fermi liquid behavior~\cite{PhysRevLett.116.076803,PhysRevB.99.195119,PhysRevX.3.031010,PhysRevX.9.021034}.

The archetypal Kondo insulator Ce$_3$Pt$_3$Bi$_4$ was comprehensively and systematically studied in the past several decades~\cite{jaime_nature,Bucher_CePtBi,Hundley_Ce343,Cooley_Ce343,takegahara_Ce343}, and regained much attention recently due to its possible connection with topological Kondo insulators~\cite{PhysRevLett.104.106408,Chang:2017aa,PhysRevLett.111.226403}; while an isoelectronic substitution of $5d$-element Pt with $4d$ element Pd,  the possible nontrivial topological properties and the metallic ground state with a large quasiparticle mass enhancement have been studied in Ce$_3$Pd$_3$Bi$_4$~\cite{Lai93,Dzsaber_ce343,Ce343_ccao,dzsaber2018giant}.
Further experiments, however, show that Ce$_3$Pd$_3$Bi$_4$ exhibits a weak activation behavior of $\Delta \sim 0.4$ meV which can be suppressed above magnetic field of $B_{\mathrm{c}}\approx$ 11 T, suggesting that the system may be very close to a metal-insulator transition point~\cite{kushwaha2019magnetic}.
Theoretically, the dynamical mean-field theory (DMFT) study revealed that the Ce$_3$Pd$_3$Bi$_4$ compound is a topological nodal line semimetal, whose metallic behaviour persists down to 4\;K, while the Ce$_3$Pt$_3$Bi$_4$ compound is a trivial Kondo insulator with an indirect gap of 6 meV below 18\;K~\cite{Ce343_ccao}.
A sharp difference between the hybridization functions of Ce$_3$Pt$_3$Bi$_4$ and Ce$_3$Pd$_3$Bi$_4$ is observed in the DFT+DMFT study~\cite{Ce343_ccao}, and in DFT study as well~\cite{Tomczak_PRB}.
Nevertheless, it has yet to be determined whether the ground state of Ce$_3$Pd$_3$Bi$_4$ compound is a small gap insulator or an exotic metal.
Although the ground state study of this compound is of significant importance, an accurate DMFT calculation at sufficiently low temperature is extremely time-consuming and unrealistic.
Here, we take an alternative approach by performing a series of calculations at different temperatures and pressure.
As a result, the magnitude of hybridization energy gap, hybridization strength, valence band edge  $E_{\mathrm{v}}$, quasiparticle weight $Z$, as well as local magnetic susceptibility can be studied as functions of temperature and pressure.
In addition,  the ground state properties can also be obtained by extrapolating temperature-dependent electronic and magnetic properties of Ce$_3$Pd$_3$Bi$_4$ to zero temperature.

\begin{figure*}
 \includegraphics[width=15cm]{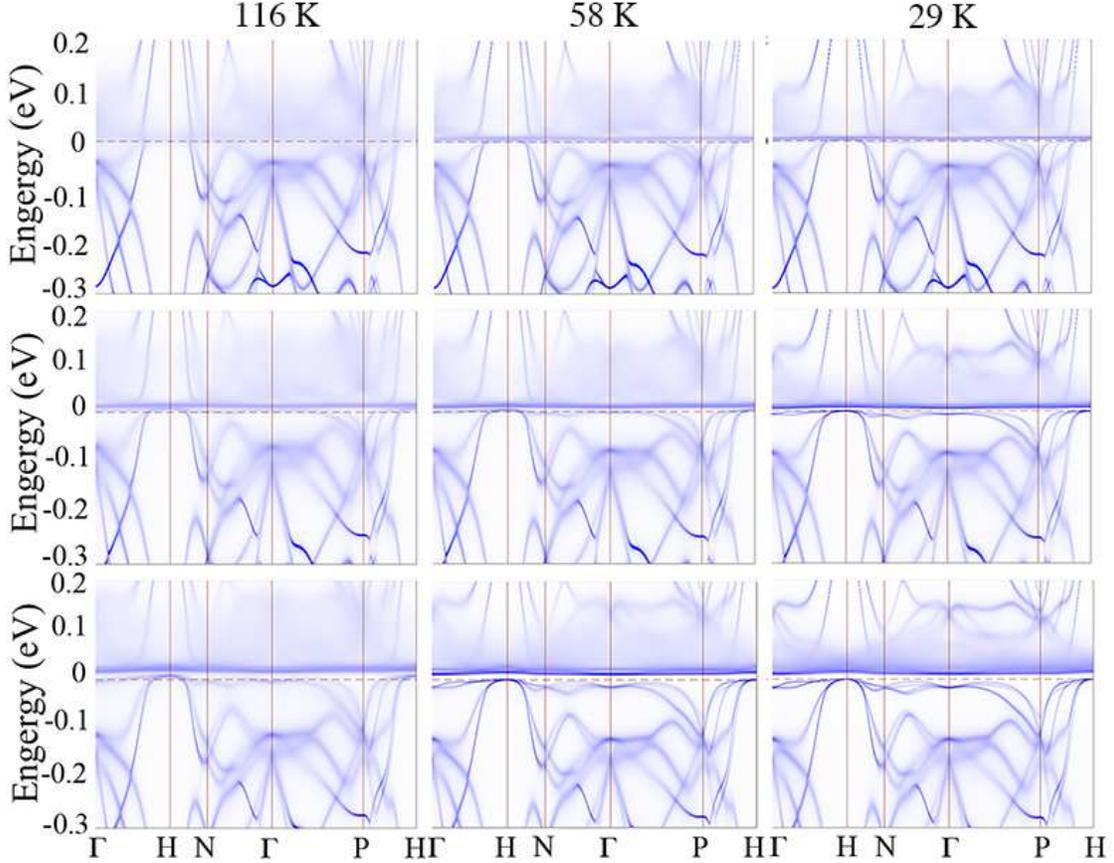}
 \caption{Momentum-resolved spectral function of Ce$_3$Pd$_3$Bi$_4$  from DFT+DMFT at 29\;K, 58\;K and 116\;K for various cell-volumes respectively.
The panels from top to bottom are 98\%, 94\% and 90\% compressed $V_{0}$, respectively.
From the top-left panel to the right-bottom panel, an energy gap around the Fermi level gradually opens under the lattice shrinking and upon cooling.
}
\label{fig:bnd}
\end{figure*}

\section{Results}
\subsection{Electronic structure of Ce$_3$Pd$_3$Bi$_4$}
The momentum-resolved spectra functions $A(\omega,\mathbf{k})$ between $E_{\mathrm{F}}-0.3$ eV to $E_{\mathrm{F}}+0.2$ eV at $T$=29\;K, 58\;K and 116\;K for 90\% $V_{0}$, 94\% $V_{0}$ and 98\% $V_0$ are shown in Figure~\ref{fig:bnd} respectively.
$V_0$ is the experimental unit cell volume at ambient pressure.
A gradual change of spectra function from incoherent to coherent behavior near the Fermi level is evident with decreasing temperature and increasing pressure (from the top-left to the right-bottom panel in Figure~\ref{fig:bnd}).
For the 98\% compressed Pd-compound (the upper panels in Figure~\ref{fig:bnd}), the spectra functions unequivocally display a blurred region in a wide temperature range from 116\;K down to 29\;K, indicative of the incoherent scattering at higher temperatures (see Supplementary Note~1 for details).
Increasing the pressure (94\% $V_{0}$ compound), the spectra become sharper near the Fermi level.
In particular, a small energy gap appears in the very vicinity of $E_{\mathrm{F}}$ at the temperature of 29\;K.
For the most compressed compound (90\% $V_{0}$) considered in this work, the energy gap can be identified at even higher temperature, implying the external pressure has driven the system away from the metallicity.
It is worth noting that the spectra of 90\% compressed Ce$_3$Pd$_3$Bi$_4$ compound at 29\;K resembles that of Ce$_3$Pt$_3$Bi$_4$ at 18\;K \cite{Ce343_ccao}, suggesting similar effect of external pressure and isoelectronic substitution by Pt, in agreement with previous study~\cite{Pickem2021}.

\begin{figure*}
 \includegraphics[width=15cm]{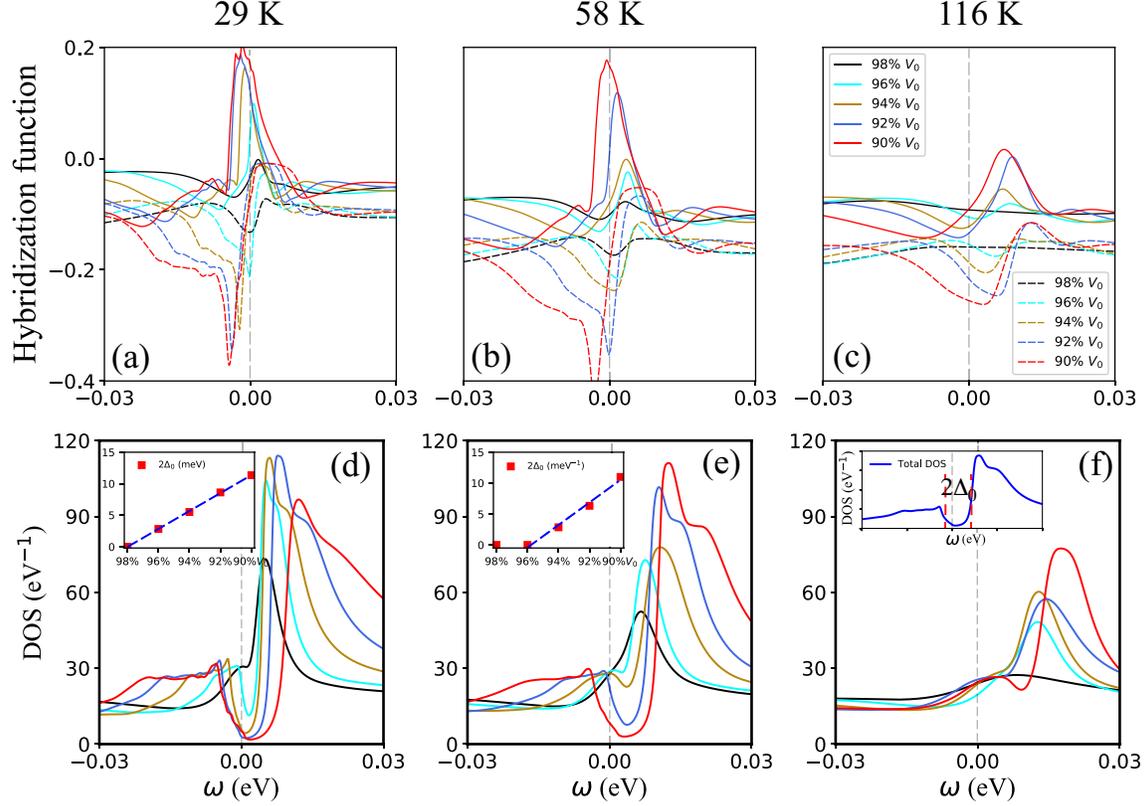}
 \caption{Pressure dependence of hybridization function and local spectra density.
(a-c) The real (solid lines) and imaginary (dashed lines) part of hybridization function.
(d-f) the integrated spectra density at 29\;K, 58\;K and 116\;K, for 98\%, 96\%, 94\%, 92\% and 90\% compressed cell-volume, respectively.
The magnitude of the energy gap is plotted as the function of cell-volume shown in the inset of d and e.
The gap-size is evaluated from the half height of the maximal intensity at the gap edge below $E_{\mathrm{F}}$ to the minimal intensity value inside the gap as schematically illustrated in the inset of (f).
The black dashed line in the inset of (f) marks the valence bands edge $E_{\mathrm{v}}$.
} 
\label{fig:hyb}
\end{figure*}

To obtain insights into the formation of energy gap upon reducing temperature and increasing external pressure, the hybridization function and integrated spectra density were illustrated in Figure~\ref{fig:hyb}.
In general the intensities of hybridization function (both the real and imaginary part) are gradually enhanced as the cell-volume decreases.
Simultaneously, the peak position is shifted away from the Fermi level, corresponding to the transfer of spectral weight from $E_{\mathrm{F}}$ to the gap edges.
This is the typical behavior of the charge gap formation due to the hybridization between the $f$-electron and conduction electrons~\cite{Bucher_CePtBi,Takeda_2006}.
The change in hybridization function is consistent with the integrated spectra density change as shown in Figure~\ref{fig:hyb}~(d-f).
The integrated spectra density at $E_{\mathrm{F}}$ initially decreases under the pressure, then begins to show a small dip for 94\% compressed Pd-compound, and finally forms a broad U-shaped gap for 90\% compressed Pd-compound at 29\;K.
Such pressure-induced behavior, however, is less obvious at 116\;K due to the collapse of energy gap at high temperature.
If we evaluate the gap size $\Delta_{0}$, as illustrated in the  insets of Figure~\ref{fig:hyb}, by the energy interval from the half-height of the maximal intensity at the gap edge below $E_{\mathrm{F}}$ to the minimum intensity value inside the gap, $\Delta_{0}$ can be well fitted with a linear function to external pressure (cell-volume) at both 29\;K and 58\;K.
Similar linear pressure dependence of energy gap was previously reported in Ce$_3$Pt$_3$Bi$_4$~\cite{Cooley_Ce343}.
In addition, for a specific pressure (volume),  $\Delta_{0}$ is also smoothly dependent on the temperature.
Hence the system continuously goes through a crossover from a metallic regime to an insulating regime upon cell-volume compressing at finite temperature.

\begin{figure*}
  \includegraphics[width=16 cm]{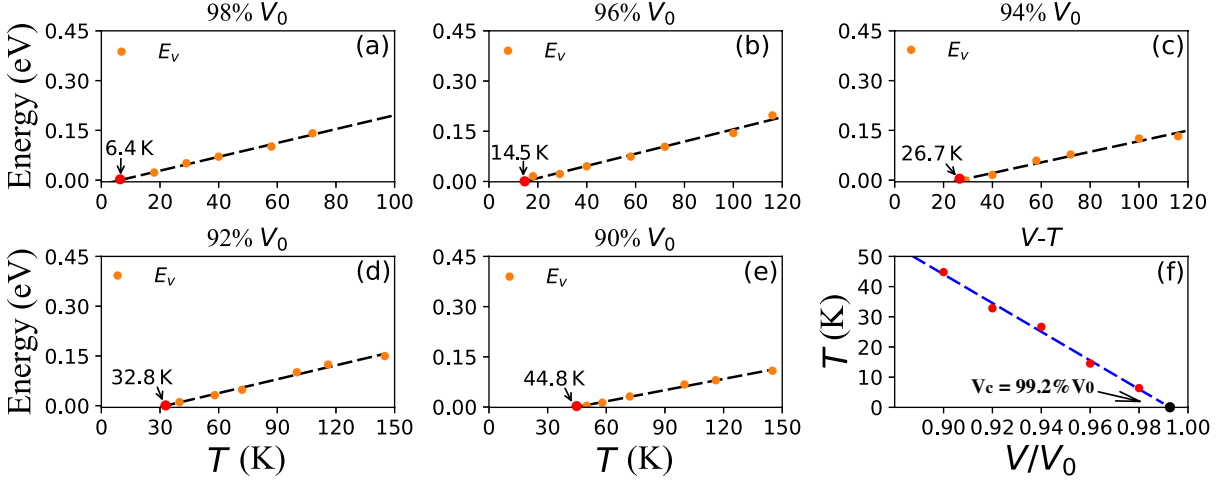}
  \caption{Temperature and pressure dependence of valence band edge.
(a-e) Temperature evolution of valence band edge $E_{\mathrm{v}}$ at 98\%, 96\%, 94\%, 92\% and 90\% of the equilibrium volume $V_{0}$, respectively.
The tops of the valence bands relative to the Fermi level are marked as blue (orange) dots.
The grey dashed lines are linear fittings to $E_{\mathrm{v}}$.
The black arrow indicates the extrapolated critical temperatures $T_1$ (marked as red dots) when the $E_{\mathrm{v}}$ cuts the Fermi level.
(f) Pressure evolution of $T_1$.
The red dots mark $T_1$ from (a-e).
The blue dashed line is the linear extrapolation to zero temperature. 
}
  \label{fig:cv_edge}
\end{figure*}

Having qualitatively established the presence of metal-insulator crossover below some certain characteristic temperature at various compressed cell-volumes,  we now address the question whether the ground state ($T \to 0$) of Ce$_3$Pd$_3$Bi$_4$ at ambient conditions  is characterized by the Kondo insulating or $f$-electron-incoherent metallic states.
As mentioned above, the spectral intensity at $E_{\mathrm{F}}$ gradually reduces upon cooling,  resulting in a red-shift of valence bands edge $E_{\mathrm{v}}$ due to the formation of energy gap.
Once the top of valence band shifts below $E_{\mathrm{F}}$, the Fermi level $E_{\mathrm{F}}$ is pinned inside the gap,  leading the system towards the insulating regime when $T>0$\;K.
Therefore, a characteristic temperature $T_1$ can be defined when $E_{\mathrm{v}}$ coincides with $E_{\mathrm{F}}$.
Below $T_1$, the electronic states around the Fermi level shall eventually become coherent, and a true gap shall appear at sufficiently low temperature.
Then we examine the shift of valence bands edge $E_{\mathrm{v}}$ relative to the Fermi level for various compressed cell-volumes in the temperature range from 18\;K to 116\;K.
The calculated results are shown in Figure~\ref{fig:cv_edge} (a-e).
By extrapolating to zero temperature, we find the valence band edge cuts the Fermi level at $T_1$ of 6.4\;K, 14.5\;K, 26.7\;K, 32.8\;K and 44.8\;K for 98\%, 96\%, 94\%, 92\% and 90\% $V_{0}$, respectively.
It indicates that the ground states are insulating at 0\;K for the corresponding cell-volume.
In Figure~\ref{fig:cv_edge} (f), we show the varying of $T_1$ as a function of cell-volume.
The $T_1$ linearly decreases as the lattice expands, which is in line with the aforementioned linear suppression of energy gap under small increasing of cell-volume~\cite{kushwaha2019magnetic}.
The temperature scale $T_1$ eventually disappears at 99.2\% $V_{0}$ (corresponding to the hydrostatic pressure of 1.1 GPa, see Supplementary Note~2), implying that the insulating phase cannot be stabilized at 0\;K when the cell-volume is larger than this value.

\subsection{Magnetic susceptibility }
To further verify this, we have also calculated the local magnetic susceptibility $\chi_{\mathrm{loc}}$ for different cell-volumes.
The results are shown in Figure~\ref{fig:chi} (also see Supplementary Note~3).
Upon lattice shrinking, the Curie-like susceptibility gradually turns to be Pauli-like with much weaker temperature dependence, indicating the reduction of local moment due to the Kondo screening.
The quantity $T^{*}$ evaluated from the maximal of $\chi_{\mathrm{loc}}$ is the characteristic temperature below which the local moment of $f$-electron and the conduction electron form an entangle state, e.g. the Kondo singlet, and the system starts entering into the coherent regime with the Fermi-liquid behavior.
As the cell-volume decreases, $T^{*}$ gradually increases, signaling the tendency of $f$-electron coherence from a low pressure phase to a high pressure phase~\cite{Marianetti_Pu,apha_beta_Ce}.
In the inset of Figure~\ref{fig:chi}, we show  $T^{*}$ as a function of cell-volume, which is qualitatively similar to the previous experimental observations~\cite{Dzsaber_ce343}.
Intriguingly, $T^{*}$ also approaches to zero at $\sim 99\% \;V_{0}$, concomitant with the disappearance of the delocalization-localization  crossover temperature, $T_1$, at the same cell-volume.
This strongly suggests that, under the ambient pressure, the  Ce-4$f$-electrons in Ce$_3$Pd$_3$Bi$_4$ compound remains localized even at zero temperature.

\begin{figure}
  \includegraphics[width=8 cm]{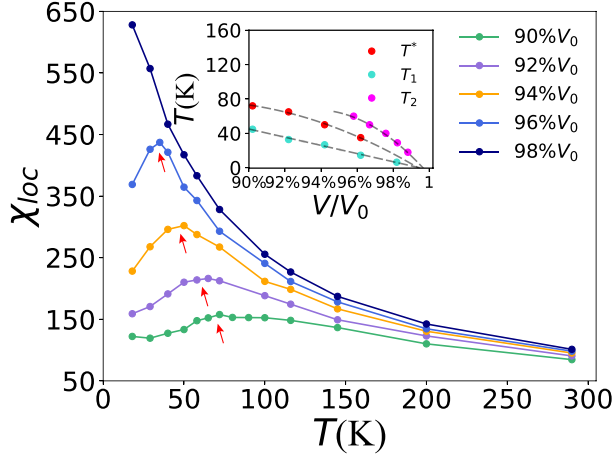}
  \caption{
Temperature dependence of local magnetic susceptibility $\chi_{\mathrm{loc}}$ for various compressed cell-volumes.
The inset shows the temperature versus pressure phase diagram.
 The characteristic temperature $T^{*}$ (marked with the red dots) associated with the formation of Kondo singlet, is corresponding to the maximums of the local magnetic susceptibility marked with red arrows in the main figure.
Correspondingly, we show the $1/\chi_{\mathrm{loc}}$ in the Supplementary Note~3.
In comparison, the pressure evolution of $T_1$ and $T_2$ from Figure~\ref{fig:cv_edge} (f) and Supplementary Figure
5(a) are illustrated in the inset as well.
The dashed line indicates the gradual suppression trend of $T^{*}$ upon lattice compression.
Note that for 98\% $V_{0}$ Pd-compound, no maximum of $\chi_{\mathrm{loc}}$ can be identified down to 18\;K.}
   \label{fig:chi}
\end{figure}  

\subsection{Upper limit of crossover temperature}
Furthermore, we can also estimate another characteristic temperature $T_2$ at which the energy gap $\Delta_0$ begins to develop around $E_{\mathrm{F}}$.
It is worth noting that the presence of such a gap at finite temperatures does not guarantee an insulating phase because the system is susceptible to the incoherent in-gap state and the alignment of the Fermi level due to thermal effect.
Therefore, it poses an upper limit of corresponding to $f$-electron localization-delocalization crossover temperature.
However, as the temperature approaches 0, the thermal effect diminishes, and the extrapolation also approaches $P_{\mathrm{c}}$ at 1.1 GPa (Figure~\ref{fig:chi} inset, as well as Supplementary Note~4).

\section{Discussion}
The following remarks and discussions are in order: Firstly, as we have stated previously, the effect of external pressure is similar to the isoelectronic substitution by Pt.
In fact, previous DFT study has shown that the radial extent of Pd(Pt)-$4d$($5d$) shall affect the hybridization strength~\citep{Tomczak_PRB}.
Therefore, the localization-delocalization transition may also be realized with Pt-substitution.
Secondly, we have observed scaling behavior in the Kondo insulating regime between unrenormalized energy gap $\Delta_0/Z$ and the maximum of hybridization function $\mathrm{Im}[\Gamma_{\mathrm{max}}(\omega)]$, defined as the peak value of imaginary part of hybridization function near the Fermi level, which suggests a critical hybridization strength of  $\mathrm{Im}[\Gamma_{\mathrm{max}}(\omega)]\approx-0.15$ eV.
This is also consistent with the critical $P_{\mathrm{c}}$ around 1 GPa (see Supplementary Note~4 for details).
Thirdly, it is informative to mention that the Hubbard $U$ has moderate influence on the critical value of the gap as well.
Since the larger value of $U$ yields more localized Ce-$4f$ moment, which is in turn harmful to a finite magnitude of energy gap,  a larger/smaller $U$ would result in a slightly larger/smaller critical pressure $P_{\mathrm{c}}$.
Nevertheless, the controversial experimental results~\cite{kushwaha2019magnetic,Dzsaber_ce343} indicate the close proximity of the ambient system close to the critical pressure, which is arguably accessible with small pressure.
It is also worth noting that similar effect of $f$-electrons localization-delocalization can also be achieved by applying magnetic field in addition to external pressure, which may serve as an explanation for the observed simultaneous vanishing of $T^*$ and $T_1$ at critical magnetic field $B_{\mathrm{c}}\approx  \;11 $ Tesla~\cite{kushwaha2019magnetic}.
Finally, at the Kondo insulating regime, we have employed topological Hamiltonian~\cite{Topological-Hamiltonian,Wang_2013} and Wilson loop method~\cite{Wilson-loop} to show that the system is topologically trivial (see Supplementary Note~5).
Since at the $f$-electron-incoherent metallic state, the system is topological nodal-line semimetals as suggested in Ref.~ \cite{Ce343_ccao}, it can be inferred that the $f$-electron localization-delocalization transition is accompanied by a topological phase transition as well.

In conclusion, we have performed a comprehensive DFT+DMFT study of pressure/temperature dependent electronic structure of Ce$_3$Pd$_3$Bi$_4$.
At ambient pressure, the ground state of Ce$_3$Pd$_3$Bi$_4$ at 0\;K is found to be metallic due to insufficient hybridization between the Ce-4$f$ electron and conduction electrons.
As the pressure increases, the hybridization enhances, and a metal-insulator transition occurs around 99\% $V_{0}$.
Using the topological Hamiltonian and Wilson loop method, the pressurized insulating phase is found to be a topologically  trivial Kondo insulator.
This leads to the observation that the metal-insulator transition is accompanied by a topological transition as well.
Our results suggest that a possible topological quantum critical point may be achieved by applying pressure in Ce$_3$Pd$_3$Bi$_4$.

\section{Methods}
To describe the electron correlation and magnetic fluctuation of Ce-compounds, the combination of density function theory (DFT) and DMFT \cite{DMFT_1996,DMFT_2006} was employed.
The DFT part was performed with the full-potential linear augmented plane-wave method implemented in the Wien2k package~\cite{schwarz_wien2k} with  generalized gradient approximation\cite{method:pbe}.
The continuous-time Monte Carlo method (CTQMC)~\cite{CT-QMC} was used as the impurity solver in its fully self-consistent version of DFT+DMFT.
The Coulomb interaction $U=6.0$ eV and Hund’s coupling $J=0.7$ eV were considered for Ce-$4f$ orbitals (see Supplementary Note~1 for details).
Since the simulated pressures are rather small, the internal atomic positions were kept the same as those under the ambient condition  when changing the lattice constant to simulate the external pressure,  The topological Hamiltonian $h_{\mathrm{t}}(\mathbf{k})=h_{0}(\mathbf{k})+\Sigma(\omega=0)$ can be obtained after the self-consistent DFT+DMFT calculations.
The zero frequency self-energy $\Sigma(0)$ was also extrapolation to zero temperature.
Both $h_{0}(\mathbf{k})$ and $\Sigma(0)$ are symmetrized with the WannSymm code \cite{ZHI2022108196}.

\section*{Data availability}
The data that support the findings of this study are available from the corresponding author upon reasonable request.

\section*{Code availability}
The latest WannSymm code is available on http://github.com/ccao/WannSymm.

\bibliographystyle{naturemag}

\section*{Acknowledgments}
The authors thank Jianhui Dai, Priscila Rosa, Qimiao Si, Huiqiu Yuan, and Wei Zhu for  discussions. The work at Los Alamos was carried out under the auspices of the U.S. Department of Energy (DOE)  National Nuclear Security Administration under Contract No. 89233218CNA000001. It was support by Pioneer and Leading Goose R \& D Program of Zhejiang 2022SDXHDX0005 (C.C), National Key R \& D Program of the MOST of China 2017YFA0303100(C.C.), NSFC 11874137 (C.C. \& C.X.),  and  the LANL LDRD Program (J.-X.Z.). J.-X.Z. was also supported by the Center for Integrated Nanotechnologies, a U.S. DOE BES user facility. The calculations were performed on the High Performance Computing Center at Hangzhou Normal University, High Performance Computing Cluster of Center of Correlated Matters at Zhejiang University, and Beijing Super-Computing Center.

\section*{Author contributions}
C.C. and J.-X.Z. initiated this work;  C.X. and C.C. performed the calculations and were responsible for the data analysis;
J.-X.Z. participated in the interpretation of numerical results.
All authors contributed to the manuscript.

\section*{Competing Interests}
The authors declare no competing interests.

\newpage
\begin{appendices}
\renewcommand{\thesubsection}{\arabic{subsection}}
\renewcommand{\figurename}{Supplementary Figure.}
\renewcommand{\tablename}{Supplementary Table.}

\section*{SUPPLEMENTARY NOTES}
\subsection{DFT+DMFT CALCULATION DETAILS}
The DFT calculation was performed based on the full-potential linear augmented-plane-wave (FP-LAPW) method as implemented in the Wien2K package~\cite{schwarz_wien2k}. The Perdew, Burke, and Ernzerhof parameterization of generalized gradient approximation to the exchange correlation functional~\cite{method:pbe} was used. Through the calculation, the number of plane waves was set to $RK_{max}=9$ and a 12$\times$12$\times$12 $\Gamma$-centered K-mesh was used to perform the Brillouin zone (BZ) integration to ensure the convergency.

The charge-self-consistent DFT+DMFT calculation was also performed within FP-LAPW basis set on a 12$\times$12$\times$12 $\Gamma$-centered K-mesh.
The continuous time Monte Carlo method (CTQMC) was applied to solve the Anderson
impurity problem with the full Coulomb interaction matrix in its rotationally invariant Slater form. We consider the Coulomb interaction $U=6.0$ eV and Hund's coupling $J=0.7$ eV
for the partially filled Ce-$4f$ orbitals. The nominal double counting scheme was considered with $n_{f}^{0}=1$ as the nominal occupancy of the ionic state. The spin-orbit coupling (SOC) was considered for the Ce-$4f$ orbitals. The crystalline electric field (CEF) splitting is one order of magnitude smaller than SOC and thus it is ignored inside the impurity solver. Hence, the local basis for
the CTQMC were projected to $|5/2\rangle$ and $|7/2\rangle$. 
To check the CEF effect,  we did a test in our previous study~\cite{Ce343_ccao} by running an additional step of DMFT self-consistency calculations with self-energy projected to distinct $m_J$ states, after the charge self-consistency was achieved.  The results suggest that the CEF is small, and its effect to spectral function is negligible even at 4 K. Since in our current study, the lowest simulated temperature is much higher, we do not expect significant CEF effect. We note that a full charge self-consistency calculation with distinct $m_J$-dependent self-energy components could also be performed, as reported in a recent work on Ce$_3$Bi$_4$Pt$_3$ by Pickem {\em et al.}\cite{Pickem2021}.
In the present work, we chose the energy range 
of hybridization from $E_{F}-10$ eV to $E_{F}+10$ eV, where the Pd-$4d$, Bi-$6p$, Ce-$4f$ and Ce-$5d$ orbitals are included. After the DFT+DFMT calculation is well converged on the  imaginary axis, we apply the maximum entropy method to do analytical continuation.
 The
spectral function is then obtained on the real axis with a 24$\times$24$\times$24 $\Gamma$-centred K-mesh. To accurately obtain the valence (conduction) band edge (Fig.~3 in main text), we employed an even more dense K-mesh with 48$\times$48$\times$48 points in the full BZ. 

\begin{figure*}
  \includegraphics[width=15 cm]{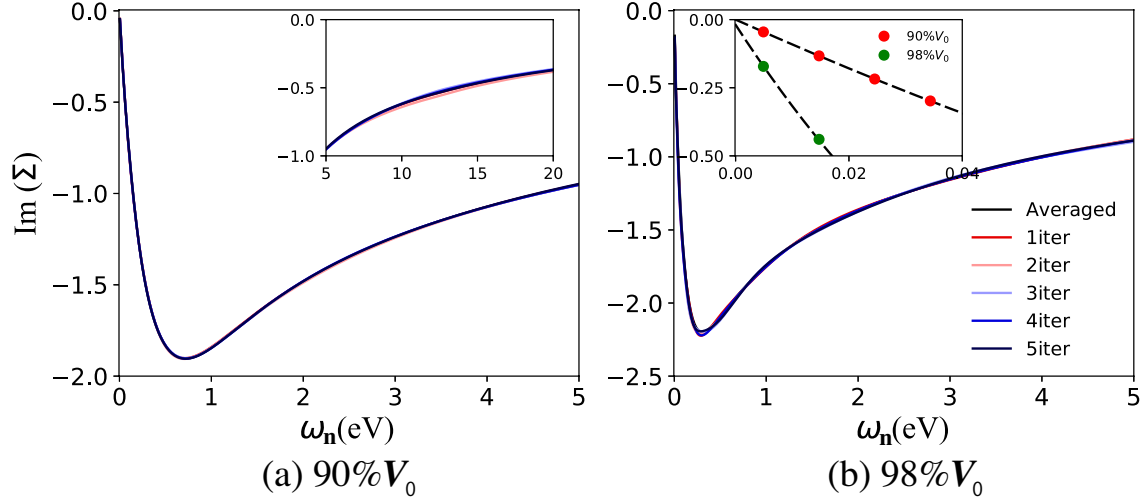}
  \caption{Imaginary part of Matsubara self-energy $\Sigma(i\omega_{n})$ of $J=5/2$
   state at 18\;K. (a) 90\% compressed cell-volume. (b) 98\% compressed cell-volume.
  Inset of (a) shows the high frequency tail of $\Sigma(i\omega_{n})$ of 90\% compressed
   compound. The low frequency
   behavior of $\Sigma(i\omega_{n})$ for both 90\% and 98\% $V_{0}$ are shown in the 
   inset of (b). The imaginary part of $\Sigma(i\omega_{n})$ is extrapolated to zero-frequency indicated by the grey dashed line. The magnitude of extrapolated value 
   $\Sigma(i\omega_{n}\to 0)$ is 21.3 meV (less than 1 meV) for 98\%$V_{0}$ (90\%$V_{0}$).
 }
   \label{fig:self-energy}
\end{figure*}

We performed the DMFT calculations for 90\%, 92\%, 94\%, 96\% and 98\% compressed cell-volume in the temperature range from 18\;K to 290\;K. Throughout the calculation, we used 96
CPU-cores and the QMC steps ranging from 2$\times$10$^8$ (for 290 K) to 8$\times$10$^8$ (for 18\;K) for each core. After the charge and energy are sufficiently converged with 25-60 DMFT iterations, we have additional 5 DMFT iterations to average the obtained self-energy in Matsubara frequency (Fig.~\ref{fig:self-energy}). As shown in
Fig.~\ref{fig:self-energy} (a-b), at the low frequency, the self-energy $\Sigma(i\omega_{n})$ of last 5 iterations and the averaged result are nearly indistinguishable, which also demonstrates the sufficient convergence of $\Sigma(i\omega_{n})$ in our calculation. When $\omega_{n} \to 0$,  the self-energy $\Sigma(i\omega_{n})$ for 90\% compressed Ce$_3$Pd$_3$Bi$_4$ approaches zero. For 98\% compressed compound, however,  $\Sigma(i\omega_{n})$ is extrapolated to a small finite value.

\begin{figure*}
  \includegraphics[width=16 cm]{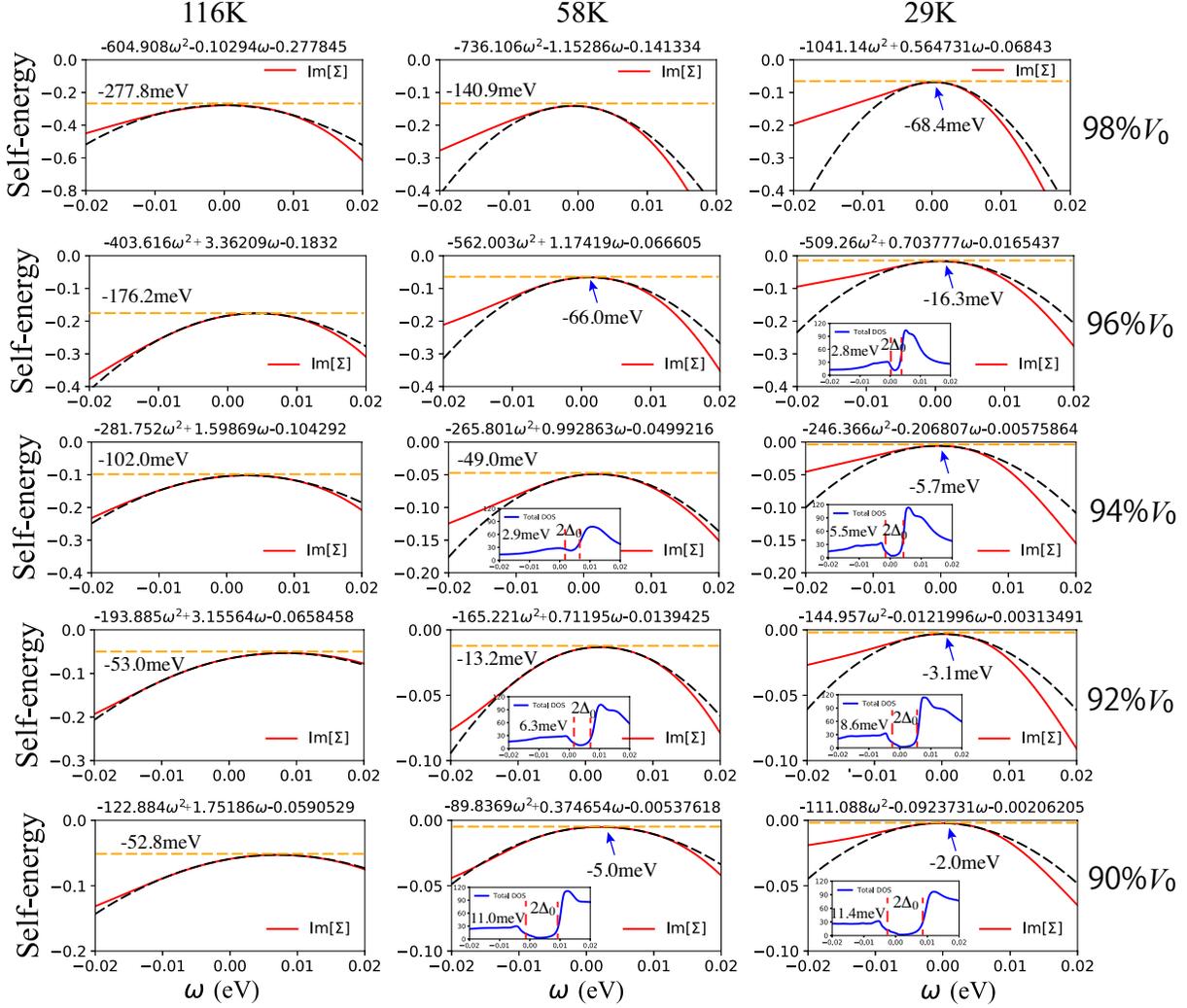}
  \caption{Imaginary part of self-energy $\Sigma(\omega)$  on real axis
  at 29\;K, 58\;K and 116\;K for various cell-volume. The red solid lines denotes the 
   imaginary part of self-energy of $J=5/2$ state,  $\Sigma_{5/2}$. 
   The dashed lines are the parabolic fitting to $\Sigma_{5/2}$.
  The orange dashed lines are guides for eye to show the $\Sigma_{0}$ from fitting. 
  For the gapped system, we show the local spectral function in the insets as well.    
 }
   \label{fig:self_energy_real}
\end{figure*}

The self-energy $\Sigma(\omega)$ on the real axis can be obtained with maximum entropy
method. We illustrate the imaginary part of Im[$\Sigma_{5/2}$] at $T$=29\;K, 58\;K and 116\;K for various cell-volume, as well as the parabolic fitting results in Fig.~\ref{fig:self_energy_real}. For the metallic system, the fitting is confined within
$[-T,T]$, where $T$ is the temperature at which the calculation is performed. While for the gapped system,
the fitting is taken inside the gap. In Fig.~\ref{fig:self_energy_real},
Im[$\Sigma_{5/2}$] can be well fitted with a parabolic function $\alpha(\omega-\omega_{0})^2+\Sigma_{0}$ in the low frequency region.  We find
the $\Sigma_{0}$ systematically varies as the function of temperature and cell-volume.
For 98\% $V_{0}$, the $\Sigma_{0}$ is substantially large in comparison with 
the gap-size, implies the incoherent non-Fermi liquid behaviour for all the temperature region on display. For 90\% $V_{0}$  at 
29\;K, the gap-size is already one order lager than $\Sigma_{0}$, indicating the coherent
Fermi liquid behaviour of current system. Such a systematic trend (from incoherent to coherent region) can also be found in the momentum-resolved spectral function in Fig.~1
in the main text.

\subsection{Pressure fitting}
\begin{figure*}
  \includegraphics[width=15 cm]{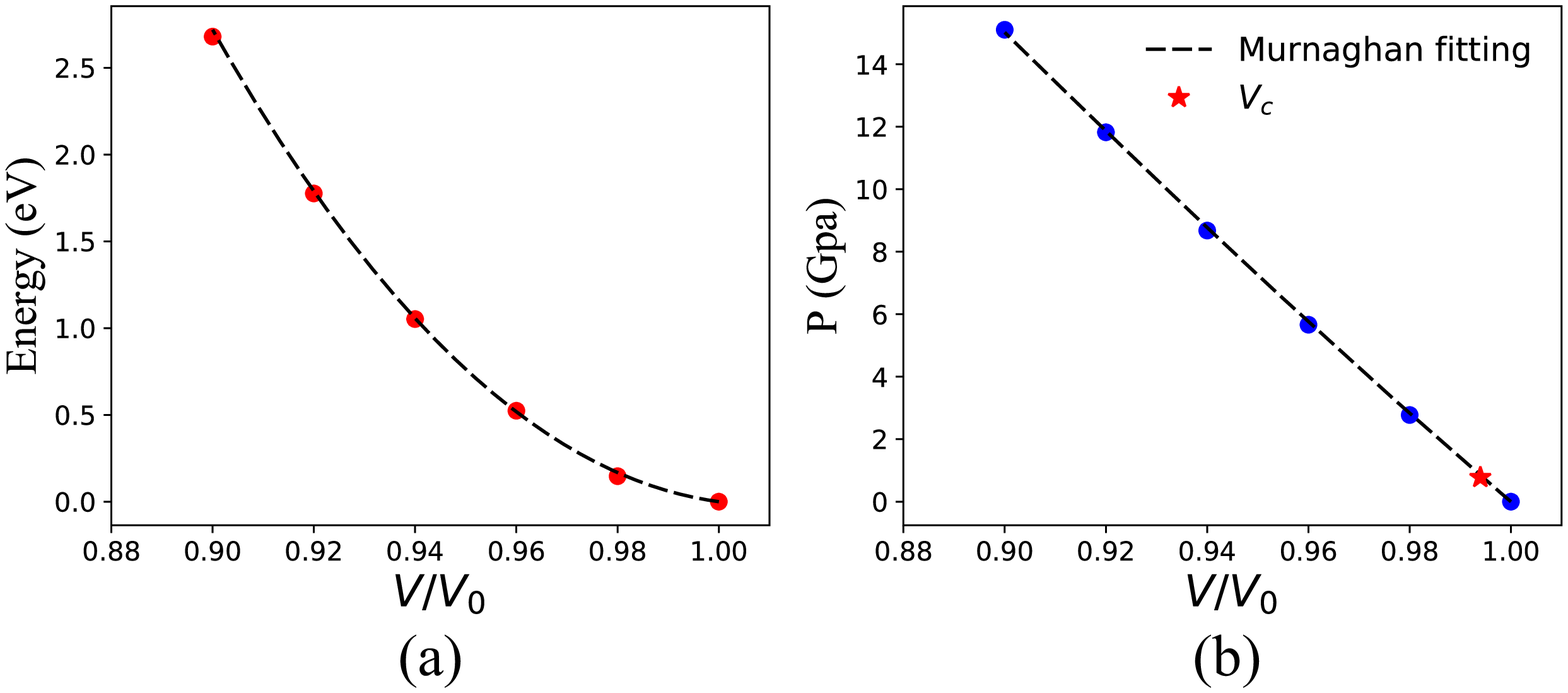}
  \caption{
   Evolution of total energy and pressure versus various cell-volume. (a) Total energy
   calculated from DFT+DMFT. The total energy of equilibrium volume
    $V_{0}$ is set
    to zero. (b) The variation of hydrostatic pressure upon lattice compressing.
    The red star denotes the critical value of cell-volume for metal-insulator transition
    discussed in the main text. The grey dashed line are Murnaghan fitting to total energy
    and hydrostatic pressure.
    }
   \label{fig:pressure_fit}
\end{figure*}

In Fig.~\ref{fig:pressure_fit} (a-b), we show the total energy and pressure
as a function of cell-volume. The total energy for various cell-volumes 
was obtained from the average value of last 5 DMFT iterations after the calculation was well converged. The pressure was fitted by using the Murnaghan equation of state, 
\begin{equation*}\nonumber
P(V)=\frac{K_{0}}{K'_{0}}\biggl{[}\biggl{(}\frac{V}{V_0}\biggr{)}^{-K'_{0}}-1\biggr{]}
\end{equation*}
where $V_0$ is the equilibrium volume, $K_{0}$ is the bulk modulus and $K'_{0}$ is its first derivative with respect to the pressure. $K_{0}$ and $K'_{0}$ were fitted from volume dependence of total energy curve. The obtained results are also cross-checked with Birch–Murnaghan fitting method. The corresponding values are listed in Table.~\ref{tab:EOS}.
The critical cell-volume $V_{c}$ (99.2\% $V_0$) are marked with red star in Fig.~\ref{fig:pressure_fit} (b). It is corresponding to the external pressure of 1.1 GPa for both fitting method.

\subsection{Inverse local magnetic susceptibility}
\begin{figure*} 
  \includegraphics[width=10 cm]{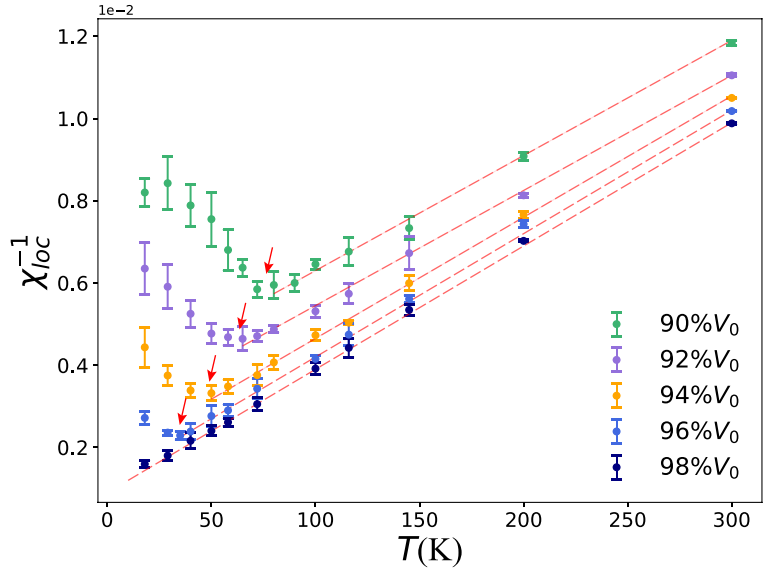}
  \caption{Temperature dependence of inverse local magnetic susceptibility for various compressed cell-volumes with error bars.
  The red dashed lines are linear fittings. The characteristic temperatures $T^{*}$ are marked with red arrow, in agree with the Fig. 4 in the main text. The error bar was evaluated from the last 5 DMFT iterations after the calculation was well converged.
    }
   \label{fig:chi_inv}
\end{figure*}

In Fig.~\ref{fig:chi_inv}, we show the temperature dependence of inverse local magnetic susceptibility $\chi_{loc}^{-1}$ for 90\%, 92\%, 94\%, 96\% and 98\%$V_{0}$ compounds.  For 98\%$V_{0}$, a linearly temperature-dependent of $\chi_{loc}^{-1}$ can be estimated even
at low temperature. As the cell volume increases, the $\chi_{loc}^{-1}$ gradually deviates from the linearly temperature dependence. The $T^{*}$ (Marked with red arrows in Fig.~\ref{fig:chi_inv}) can also be roughly estimated as the temperature when the $\chi_{loc}^{-1}$ begins to deviate from the linearity.

\subsection{Pressure and temperature dependent local spectra density, gap, renormalized factor and hybridization function}

\begin{figure*}
\includegraphics[width=15 cm]{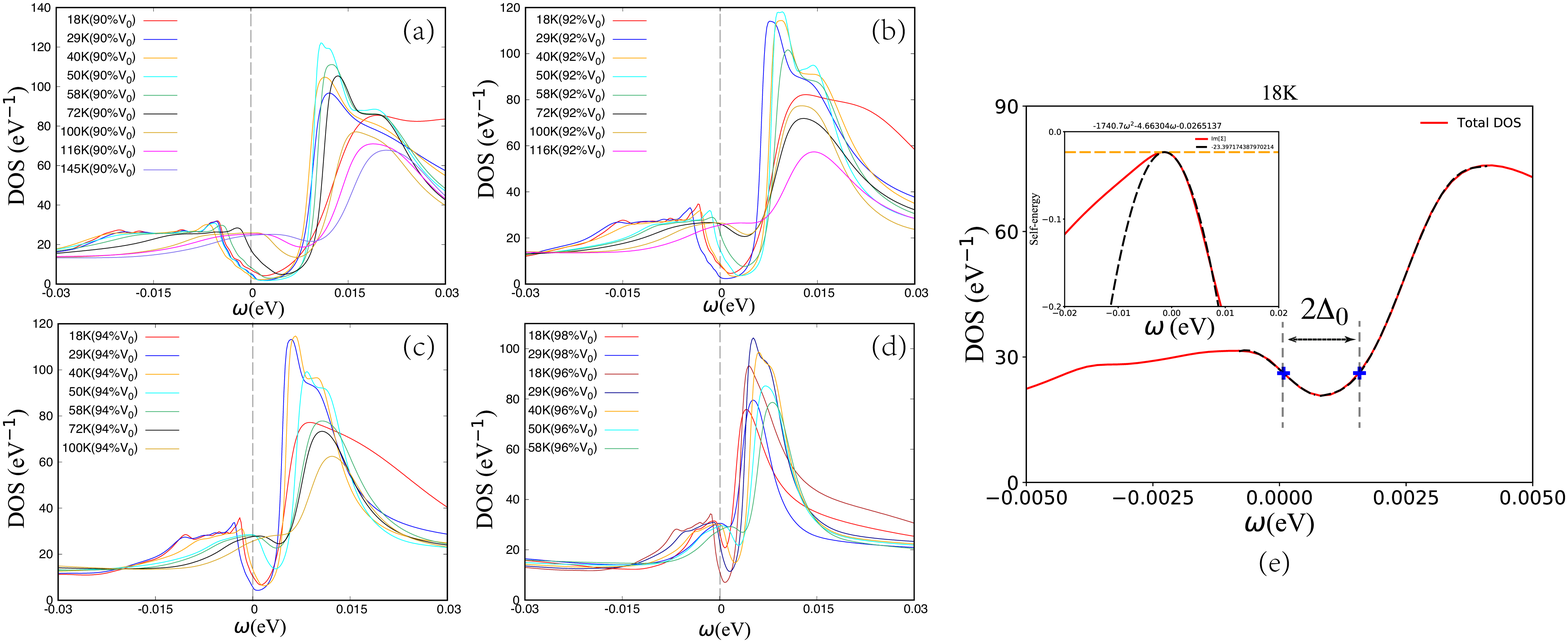}
\caption{
 Pressure and temperature dependent of local spectra density.
 (a-d) The local spectra density plot for 90\%$V_{0}$, 92\%$V_{0}$, 94\%$V_{0}$,
 96\%$V_{0}$ and 98\%$V_{0}$ compressed compounds at various temperatures. The gap 
 size are evaluated from the half height of the maximal intensity at the gap edge below $E_{F}$ to the minimal intensity value inside the gap as mentioned in the main text.
 (e) The local spectra density 98\%$V_{0}$ compressed cell-volume. The imaginary part of Im[$\Sigma_{5/2}$] on real-frequency axis for the current system is illustrated in the inset of (e). The black dashed line are the least-square fitting of local spectra density and Im[$\Sigma_{5/2}$] within the gap.
 }
  \label{fig:dos}
\end{figure*}

\begin{figure*}
\includegraphics[width=15 cm]{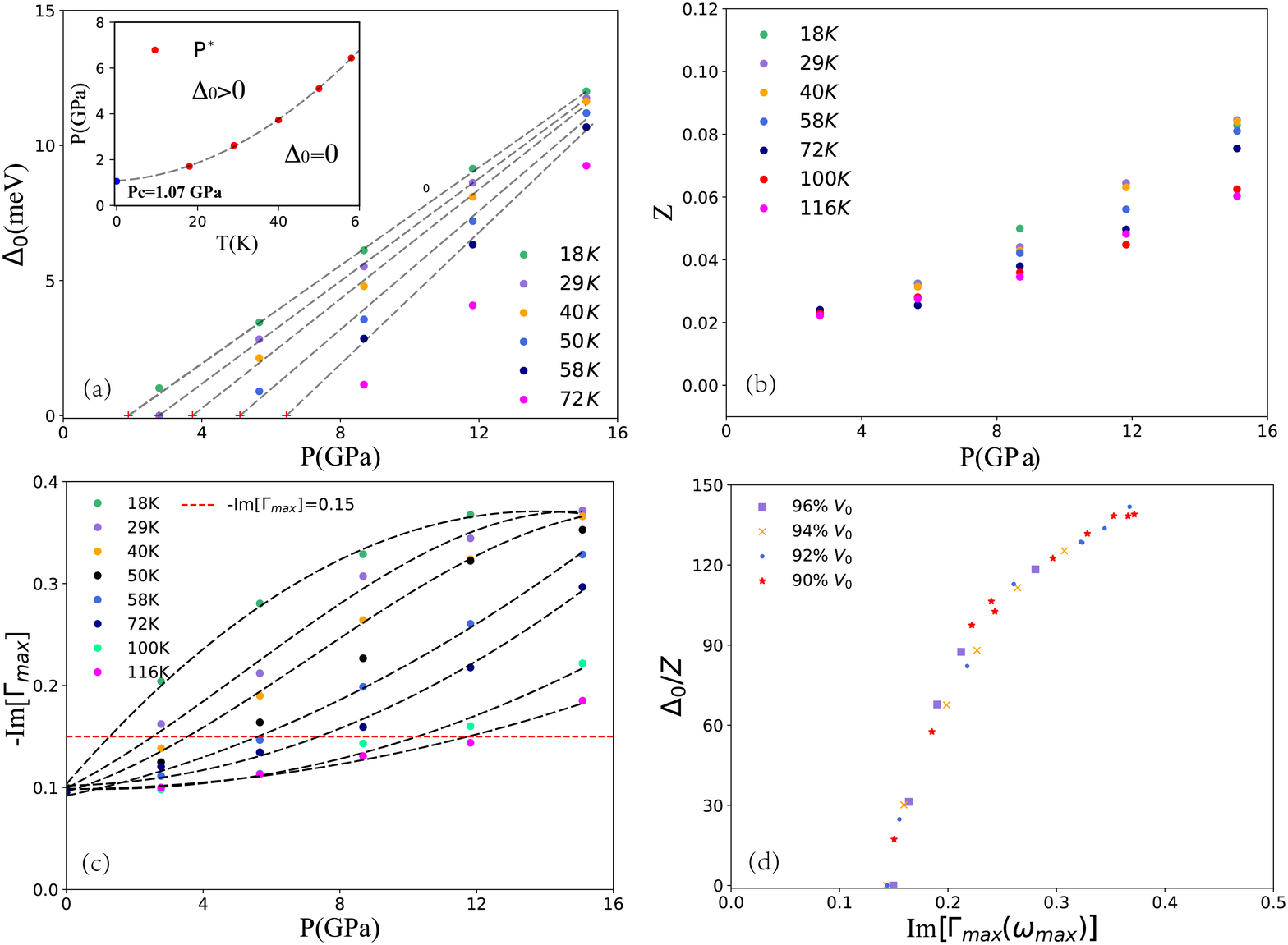}
 \caption{Temperature and pressure dependent gap $\Delta_{0}$, renormalized factor $Z$ and hybridization function. (a) The pressure dependence of gap $\Delta_{0}$. $\Delta_{0}$ was linearly fitted as a function of pressure in the temperature range from 18 K to 58 K .  The characteristic $P^{*}$ at which $\Delta_{0}=0$ were marked with red plus. In the inset, $P^{*}$  was fitted as a function of temperatures marked with dashed line.  (b) The temperature and pressure dependence of quasiparticle weight $Z$ ($m^{*}/m=Z^{-1}=1-\frac{\partial Re[\Sigma(\omega)]}{\partial \omega}$). (c) The imaginary part of hybridization function -Im[$\Gamma(\omega)$] at its peak value vs pressure from 18\;K to 116\;K. $\omega_{max}$ is the corresponding frequency, where -Im[$\Gamma(\omega)$] reaches the maximal amplitude -Im[$\Gamma_{max}(\omega)$]. The values of -Im[$\Gamma_{max}(\omega)$] at $P=0$ were from our previous calculation \cite{Ce343_ccao}. The red dashed line marks -Im[$\Gamma_{max}(\omega)$]=0.15 eV, where the $\Delta_{0}$ disappears as shown in (d). (d) The unrenormalized gap $\Delta_{0}/Z$ vs -Im[$\Gamma_{max}(\omega)$] for various compressed cell-volume. $\Delta_{0}/Z$ vanishes at -Im[$\Gamma_{max}(\omega)$] $\approx 0.15$ eV. }
   \label{fig:physquantity}
\end{figure*}

We illustrate the evolution of gap size for all the compressed
cell-volume from high temperature (gap opening) to low temperature in Fig.~\ref{fig:dos} (a-d). 
It is worth noting that only at zero temperature the metal-insulator phase transition is 
well defined. At finite temperature, due to the incoherent scattering (the nonzero value of the imaginary part of self-energy), there is only a crossover region between the metallic region and insulating region.
As shown in Fig.~\ref{fig:dos} (e), although the gap   already opens at 18\;K 
for 98\%  $V_0$, the system is still in the crossover region instead of the insulating region, since the imaginary part of self-energy ($\Sigma_{0} \thicksim $ 26.5 meV, defined in Fig.~\ref{fig:self_energy_real} ) is much larger than the gap size $2\Delta_{0} \thicksim$ 2 meV.

The pressure evolution of  gap size $\Delta_0$ is shown in Fig.~\ref{fig:physquantity} (a).  In general, $\Delta_0$ displays linear pressure dependence within the low temperature region ($T \leq 58$ K), similar to the observation in Ce$_3$Bi$_4$Pt$_3$ experimentally~\cite{Cooley_Ce343}. When the temperature further increases, the $\Delta_0$-$P$  curve deviates away from the linear behavior. This is possibly due to the thermal effect.  By extrapolating to $\Delta_{0}=0$, we define the gap opening pressure $P^{*}$,  below which the system can be in the metallic regime or the crossover regime, rather than in the insulating phase. Therefore, the characteristic $P^{*}$ is the lower limit pressure, where the system begins to enter  the insulating regime from the metallic side when $P$ increases at a specific temperature.  Correspondingly, for a certain pressure, this specific temperature is the upper limit temperature $T_{2}$ mentioned in the main text. In the inset of Fig.~\ref{fig:physquantity} (a), $P^{*}$ was fitted parabolically as function of $T$. When $T$ approaches 0K, the thermal effect diminishes and $P^*$ should also approach the critical pressure, $P_c$, at 0 K. Our fitting yields $P^*=1.07$ GPa at 0 K. In the main text, we show the corresponding variation of characteristic $T_{2}$  as a function of cell-volume in the inset of Fig. 4.

Figure~\ref{fig:physquantity} (b) shows the temperature and pressure dependence of quasiparticle weight $Z$. In Fig.~\ref{fig:physquantity} (c), the maximal amplitude of hybridization function $-\text{Im}[\Gamma_{max}(\omega)$] at $\omega_{max}$ around the Fermi level was plotted as the function of pressure at $T$=18\;K, 29\;K, 40\;K, 50\;K, 58\;K, 72\;K, 100\;K and 116\;K.  The values of $-\text{Im}[\Gamma_{max}(\omega)$] at ambient pressure were obtained from 
Ref.~\cite{Ce343_ccao}. The pressure dependence of $-\text{Im}[\Gamma_{max}(\omega)$] evolves from concave tendency at high temperature to convex tendency at low temperature. In Fig.~\ref{fig:physquantity} (d),  we show that the unrenormalized gap ($\Delta_{0}/Z$) size continuously approaches zero as the maximal amplitude of $-\text{Im}[\Gamma_{max}(\omega)$] decreases for various compressed cell-volume.  The critical value of $-\text{Im}[\Gamma_{max}(\omega)$] $\approx$ 0.15 eV, where the gap  disappears. In Fig.~\ref{fig:physquantity} (c), the critical value of $-\text{Im}[\Gamma_{max}(\omega)$] was marked with red dashed line, below which the gap vanishes. The corresponding critical pressure $P$ for the  disappearance of gap at 18\;K is around 1.26 GPa. As the temperature further decreases, $P$ decreases as well. When the temperature approaches zero, $P$ is expected in the range between 0 to 1.26 GPa, where
the metal-insulator transition occurs.

\begin{table*}
\caption{List of total energy $E_0$ and pressure $P$ for 90\%, 92\%, 94\%, 96\%
and 98\% $V_{0}$ at 29\;K. The total energy of equilibrium volume $V_{0}$ is set
to zero.The values inside (outside) the bracket were fitted with Murnaghan equation
of state (Birch–Murnaghan equation of state).
\label{tab:EOS}}
\begin{tabular}{c| cccc|cccc|cccc|cccc|cccc|cccc}
  \hline
              &\multicolumn{4}{c }{\quad 90\%$V_{0}$\quad}
              &\multicolumn{4}{c }{\quad 92\%$V_{0}$\quad}
              &\multicolumn{4}{c }{\quad 94\%$V_{0}$\quad}
              &\multicolumn{4}{c }{\quad 96\%$V_{0}$\quad}
              &\multicolumn{4}{c }{\quad 98\%$V_{0}$\quad}
              &\multicolumn{4}{c }{\quad $V_{0}$ \quad}\\
 \hline
 \quad$E_{0}$ (eV) \quad &\multicolumn{4}{c }{2.680 (2.680)}
              &\multicolumn{4}{c }{1.776 (1.775)}
              &\multicolumn{4}{c }{1.053 (1.053)}
              &\multicolumn{4}{c }{0.524 (0.524)}
              &\multicolumn{4}{c }{0.147 (0.147)}
              &\multicolumn{4}{c }{0.000 (0.000)}\\    
\hline
 $P$ (GPa)    &\multicolumn{4}{c }{15.10 (15.02)}
              &\multicolumn{4}{c }{11.82 (11.87)}
              &\multicolumn{4}{c }{8.68  (8.77)}  
              &\multicolumn{4}{c }{5.66  (5.76)}
              &\multicolumn{4}{c }{2.77  (2.83)}
              &\multicolumn{4}{c }{0.00  (0.00)} \\
\hline
\end{tabular}
\end{table*}

\begin{figure*}
  \includegraphics[width=15 cm]{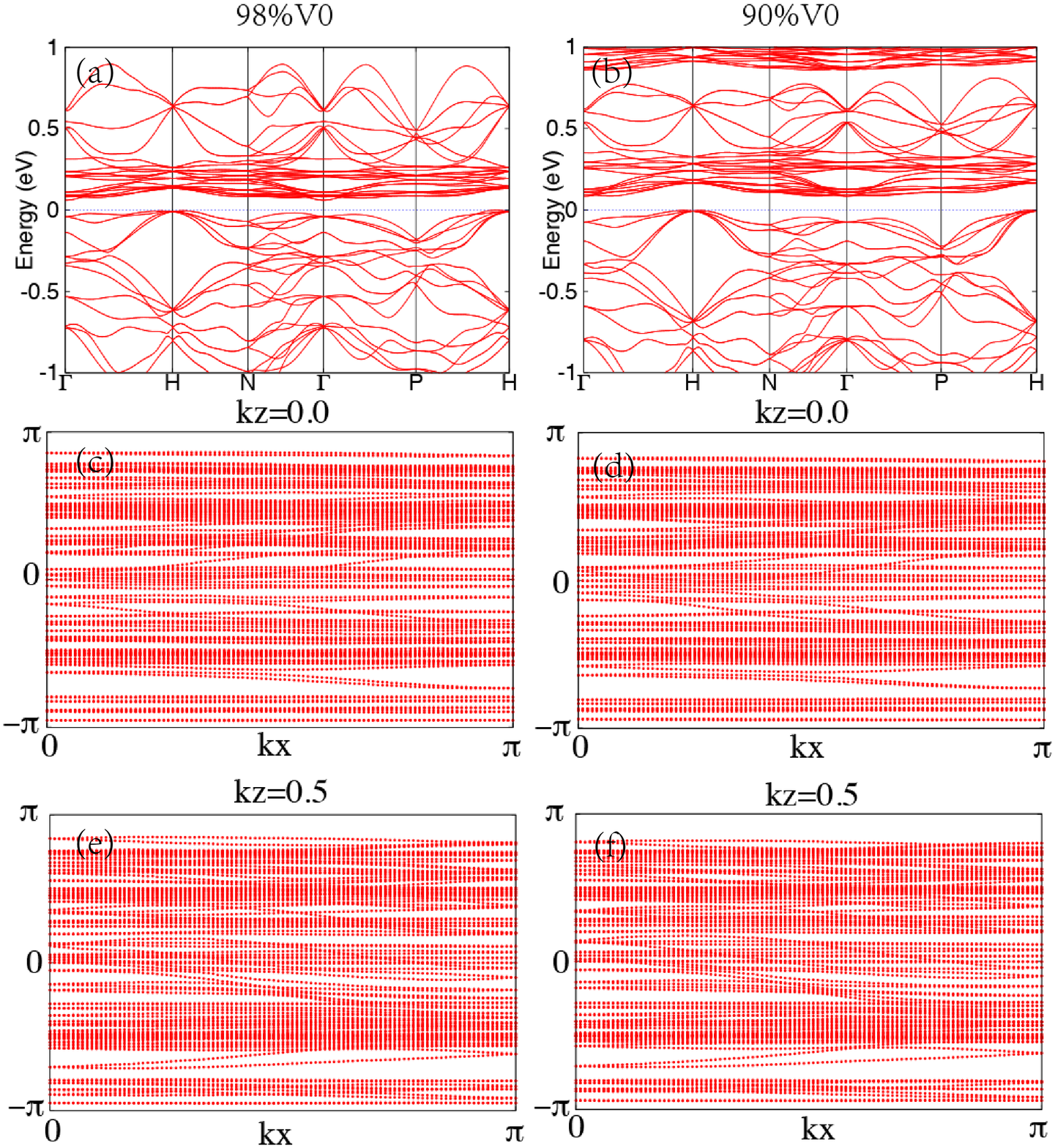}
   \caption{Energy spectra of topological Hamiltonian and
   the evolution of Wannier charge center. (a-b) The spectra 
   of topological Hamiltonian $H_{t}=H_{0}+\Sigma(\omega=0)$
   for 98\%$V_{0}$ and 90\%$V_{0}$, respectively, where 
   $\Sigma(\omega=0)$ is the zero frequency self-energy extrapolated 
   to zero temperature. (c-f) The corresponding evolution of Wannier 
   charge center for the occupied states at $\mathbf{k}_{z}$=0.0 and 
   $\pi$ plane.
   }
   \label{fig:wanner_center}
\end{figure*}

\subsection{ Topological property}
The topological property of current system can be described with
topological Hamiltonian $h_{t}(\mathbf{k})=h_{0}(\mathbf{k})+\Sigma(0,\mathbf{k})$,
which is encoded in zero frequency~\cite{Topological-Hamiltonian,Wang_2013}.
By extrapolating to zero temperature, we can obtain the zero frequency
self-energy $\Sigma(0,\mathbf{k})$ when $T \to 0$. In Fig.~\ref{fig:wanner_center} (a-b), we show the energy spectra of
topological Hamiltonian for 98\% $V_{0}$ and 90\% $V_{0}$, respectively. It  
is worth mentioning that the energy spectra of $h_{t}$ is not exactly
the quasiparticle excitation spectra but the topological property of
interacting system is correctly captured by $h_{t}$~\cite{Wang_2013}.
The evolution of Wannier charge center was calculated with the aforementioned topological Hamiltonian $h_{t}$ for all the occupied 
states with Wilson-loop approach \cite{Wilson-loop}. In either
98\% $V_{0}$ or 90\% $V_{0}$ system, the evolution pattern is
trivial, indicating the normal insulating behavior of current system.

\end{appendices}

\end{document}